# On the excitation of Ion Acoustic Soliton in quiescent plasma confined by multi-pole line cusp magnetic field

Zubin Shaikh[1,3], A.D.Patel[1], Meenakshee Sharma[1], H.H.Joshi[3]

N.Ramasubramanian[1,2]

1. Institute for Plasma Research, Gandhinagar-382428, India
2. Homi Bhabha National Institute, Anushaktinagar-400094, Mumbai, India
3. Department of Physics, Saurashtra University, Rajkot-360005, India

*zubin.ipr@gmail.com*

## ABSTRACT

This paper presents the detailed study of the controlled experimental observation and characterization of Ion Acoustic soliton in the quiescent argon plasma produced by filamentary discharge and confined in a multi-pole line cusp magnetic field device named Multi-pole line Cusp Plasma Device (MPD). In this system, the electrostatic fluctuations are found to be less than 1% ($\delta I_{isat}/I_{isat} < 1\%$), a characteristic of quiescent plasma. Ion acoustic soliton has been excited in MPD, and its propagation velocity and width of them are measured experimentally and compared with the 1-D Korteweg-de Vries (KdV) equation. The interaction of two counter-propagating solitons is also investigated to confirm propagation's solitary nature further. After the successful characterization of ion-acoustic soliton, the effect of varying the cusp magnetic field on the propagation of ion-acoustic soliton has been studied. It is experimentally observed in MPD that the pole cusp magnetic field value influences the excitation and propagation of solitons. The soliton amplitude increases with the pole field up to some value $B_p \sim 0.6 kG$, then decreases with the further increase in field values. Meanwhile, the width of the soliton shows different behavior. The role of primary electron confinement by cusp magnetic field geometry has been used to explain the observed results.

## 1. Introduction

Plasma is a complex fluid that supports various kinds of waves. High frequency ($\omega \geq \omega_{pe}$) and low frequency ($\omega \geq \omega_{pi}$), electromagnetic and electrostatic, Linear and Non-linear waves can propagate in plasma[1]. An ion-acoustic soliton is a typical non-linear electrostatic wave that widely appears in laboratories[2,3,4] space, and astrophysical plasma[5,6]. The non-linear behavior of this wave is a matter of interest and plays a critical role in many fundamental processes of plasma physics[7,8,9,10]. There are several areas where solitons are observed, like, Water[11], plasma[2], liquid helium[12], and optics[13] and upper ionosphere and lower magnetosphere[5,14].

Not only in fundamental plasma physics but also due to its non-linear properties, a soliton is crucial in other research areas. Recently Hideki et al.[15] proposed a new acceleration mechanism for charged particles by Ion acoustic solitons in plasma. Niemann et al.[16] experimentally observed the parametric two-ion decay instability of ion-acoustic waves driven by a laser beam in a laser-produced plasma. Many experimental works have been carried out for the excitation[17,18], propagation[19,20] collisions[21], etc., for ion-acoustic soliton. To investigate its characteristic, various authors have theoretically[22,23], and numerically[24,25,26] derived the KdV equations. KdV equation shows the dependency of the width and velocity of solitons upon their amplitude. Apart from that, it is well established from numerical solutions of the KdV equation that soliton regain its original identity after interacting with each other[2,27,28,29]. Authors have studied the collisions of solitons in electron ion plasma[30,31,32], plasmas with positive and negative ions with Boltzmann electrons[33], and plasma with electron ions and positrons[34,35,36].

H. Ikezi[2] experimentally discovered and characterized the ion-acoustic soliton for the first time in a unique device called the double plasma device[37]. Watanabe[38,39] peruses the method of Ikezi to excite soliton in plasma by using a floating wire grid placed inside the plasma. The experimentally measured the width and velocity matched with a solitary wave solution acquired by Sakanaka[40]. Nishikawa and Kaw[41] have obtained the WKB solution for propagating the ion-acoustic solitary wave in plasma with a density gradient. They found that soliton's properties are modified as it propagates towards the higher density region. John & Saxena[42] experimentally verified it in a large double plasma machine.

Jacqueline Hill[43] has studied the propagation of ion-acoustic soliton perpendicular to a magnetic field (transverse field), while Raychaudhuri[44] has studied it in a longitudinal field in the same double plasma device[45]. A Sharma[46] has numerically studied the magnetic field effects on the electromagnetic solitons, while Lin Wei[47] has studied the effects of the bounded geometry on the ion acoustic wave.

In the present work, we study excitation, propagation, and characterization of ion-acoustic soliton in plasma confined by a multi-pole line cusp magnetic field and the impact of varying cusp magnetic field on the modification of properties of ion-acoustic soliton. The layout of the paper is as under. Section 1.1 describes the basic KdV equation and conditions for soliton.

Section 2 describes the details of the experimental setup and wave excitation and detection technique. Section 3 describes the detailed experimental results and discussions. Section 4 describes the effect of the cusp magnetic field on soliton, followed by a summary in section 5.

## 1.1 Introduction KdV Equation and condition for soliton[2]

The asymptotic solution of KdV equation[2,27,29] describes the finite-amplitude ion wave in plasma. The electron density profile in the solitary wave pulse is given by

$$\frac{\partial \psi}{\partial \xi} + \frac{\partial \psi}{\partial \tau} - \psi \frac{\partial \psi}{\partial \tau} - \frac{1}{2}\frac{\partial^3 \psi}{\partial \tau^3} = 0 \tag{1}$$

The stationary state solution of it gives the electron density profile in the solitary wave pulse.

$$\tilde{\eta} = \delta n \, sech^2 \left[\frac{x - ut}{D}\right] \tag{2}$$

where,

$$u/C_s = 1 + \frac{1}{3}\delta n/n_o \tag{3}$$

and,

$$\left(D/\lambda_D\right)^2 = 6n_o/\delta n \tag{4}$$

Where $\psi$ is perturbed to unperturbed plasma density *(n/n$_o$)*

$\xi = x/\lambda_D$ is the spatial co-ordinate normalized by Debye Length $\lambda_D$

$\tau = \omega_{pi} \cdot t$ Time normalized by the ion plasma frequency $\omega_{pi}$

$n_o$ is the unperturbed plasma density

$C_s$ is the Ion-Acoustic speed

$\delta n$ is the amplitude of the soliton.

From these KdV equations, two things can be understood physically.
1. As the soliton amplitude increase, the width of the soliton decreases, and reported results of KdV suggests that $\delta n D^2 \sim const$. I.e., The Square of width times the amplitude is constant

2. Velocity of the soliton is observed to be more than the ion acoustic speed. It implies that the Mach number must be more than unity or one ($u/C_s > 1$).

The solitons characterization needs these two basic necessary conditions. The above two relations are definite characteristics of KdV solitons which are often used to identify solitary waves observed experimentally. Apart from these two mentioned characteristics, one intriguing phenomenon is observed when two counter-propagating solitons collide with each other. During the interaction, both solitons merge and generate a single soliton. After the merging, they separate from each other without losing their identity.

## 2. Experimental Setup

The experiment is carried out in Multi-cusp Plasma Device (MPD)[48,49]. MPD is a unique device that is able to produce different multi-pole cusp magnetic field configurations with different pole magnetic field strength[50]. The MPD consists of a cylindrical vacuum vessel made of stainless steel having a wall thickness of 6mm, 1500 mm long, and a diameter of 400mm. The chamber is evacuated by a Turbo Molecular Pump (TMP) backed by the rotary pump through a conical reducer at one side of the chamber. Base pressure of $1\times10^{-6}$ mbar was achieved and measured at the center of the device by a hot ionization gauge. MPD consists of six rectangular-shaped electromagnets with Vacoflux-50 as core material for producing the variable multi-pole line cusp magnetic field. These electromagnets are placed on the periphery of the vacuum vessel, and each magnet is placed 60 degrees apart.

The direction of the current in these electromagnets can be altered to produce different magnetic field geometry. The experiments were performed with 12 pole cusp magnetic field configuration. For producing 12 pole cusp configuration current in all six magnets are in the same direction; hence all six magnets will produce one type of pole, and another virtual pole will be produced in between two magnets; hence a total of 12 cusps will be there of six dipoles. Compared to MPD's possible magnetic field configuration, the 12-pole cusp configuration has a more uniform field-free region (~20cm). Plasma confined in this uniform field-free region, all the basic plasma parameters (electron temperature ($T_e$), plasma density ($n_e$), plasma potential ($V_p$), floating potential ($V_f$), and fluctuations $\delta I_{isat}/I_{isat}$ are uniform. Which is suitable for wave-particle and interaction studies[50–52].

The filamentary Argon discharge plasma was produced using a hot filament-based cathode source. The plasma source (cathode) is two dimensional (8cm x 8cm) vertical array of five tungsten filaments; each filament has a 0.5 mm diameter and 8 cm length. They are built in the conical reducer such that the filaments are inside the main chamber, where the magnetic field is low. Also, it has been taken care to push the source well inside the main chamber to avoid the edge effects of the magnets. These filaments are powered by a 500 A, 15 V floating power supply, usually operated at around 16 - 19 A per filament. The chamber was filled with Argon gas through a needle valve to a working pressure of 8 x $10^{-5\,m}$ Bar. The source is biased with a voltage of - 50 V with respect to the grounded chamber walls using a discharge power supply. The primary electrons (hot electrons) emitted from the filaments are bound to travel in the electric field directions, i.e., towards the chamber wall; during this path, they collide with the neutral argon gas. These collisions ionized the gas atoms. These ions and electrons are confined in a multi-pole line cusp magnetic field. With this configuration, a discharge current

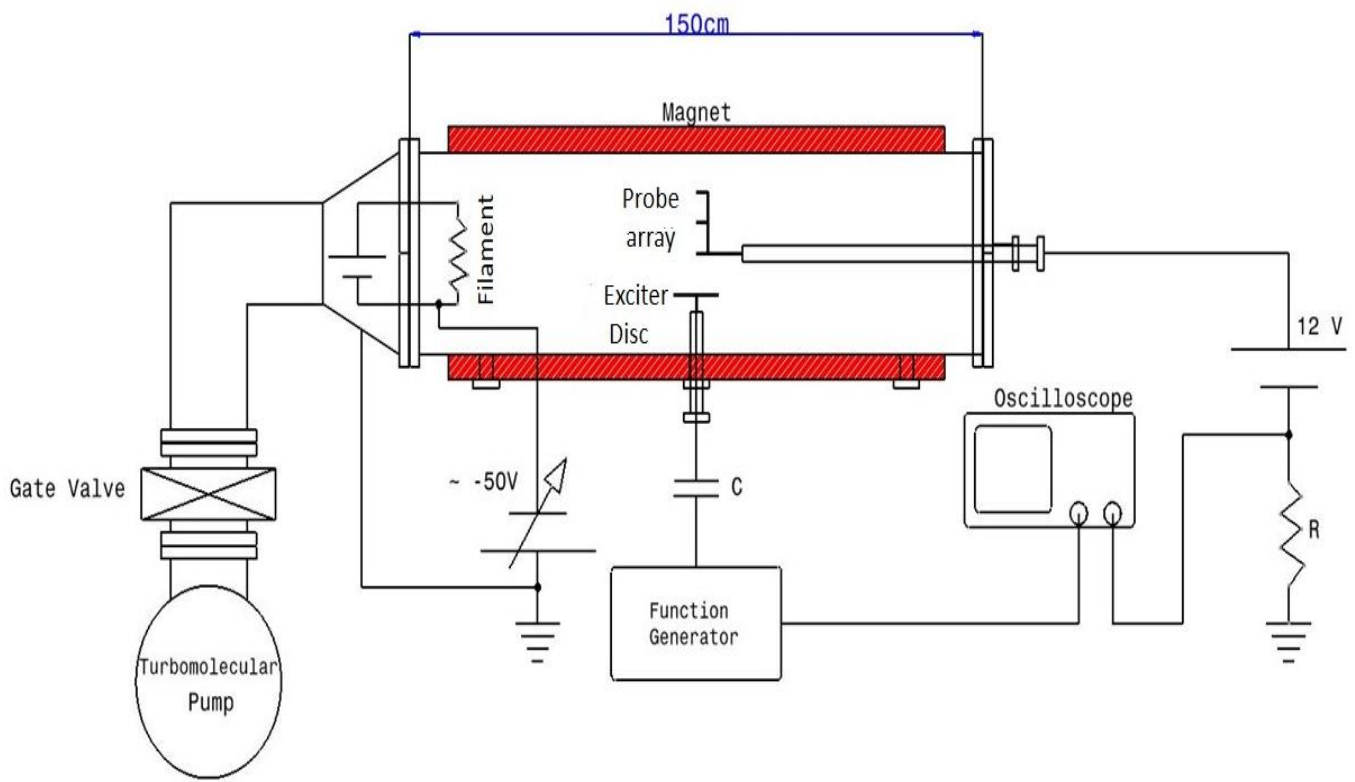

of ~ 3-6A is achieved. Figure 1 shows the schematic diagram of the device.

*Figure 1: Schematic diagram of the Multi-Cusp Plasma Device (MPD) Experimental setup*

Mean plasma parameters are measured using the I-V characteristics of a Single Langmuir Probe (SLP). Typical measured plasma parameters are at the midplane of device are: Plasma Density $(n_e)$ ~ $10^{16}$ m$^{-3}$, electron temperature $(T_e)$ ~ 3-4 eV, Plasma potential $(V_p)$ ~ 4-5 V, Ion Temperature $(T_i)$ ~ 0.4 eV $(T_i = T_e/10)$, Ion plasma frequency $(\omega_{pi})$ ~ 4 x$10^6$ Hz for - 50V discharge voltage and 8x10$^{-5}$ mbar working pressure and $B_p = 0.6kG$ (Magnet Current $I_{mag} = 80A$). The experiment report below was performed with these fixed parameters until unless specified.

## 2.1 Wave excitation and detection techniques

A relatively large amplitude perturbation voltage to the floating exciter metal disk through a capacitor has been applied to excite the solitons in plasma, which is the compressive wave of plasma density. By this mechanism, the compressive pulse and a rarefactive pulse that follows the compressive pulse are also excited, and the grid itself does not emit the plasma[53,54]. The exciter disk is placed inside the plasma at the central plane of the device, where the magnetic field is minimum, and the plasma is uniform and quiescent[50]. This exciter grid is a solid disk of molybdenum with a diameter of 50mm and a thickness of 0.25mm. The solid disk has been used because of the finite sheath thickness around it[53,55].

A single pulse sinusoidal voltage of ~$20V_{pp}$ at $90kH_z$ frequency has been applied to the grid to excite the soliton in plasma. The selection of perturbation frequency also satisfies the range of $\omega/\omega pi < 0.7$[56]. A PA-85 amplifier-based circuit with 10 gain has been made to amplify the perturbation signal of 90 kHz of frequency. This circuit has increased only the amplitude while the frequency remains unchanged.

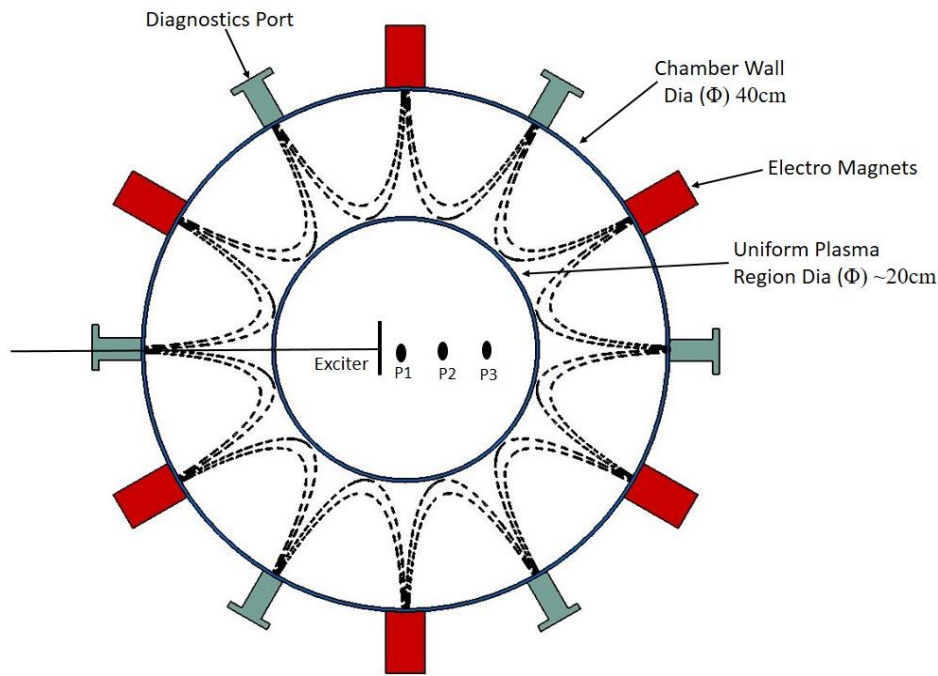

*Figure 2: Cross Section view of the experimental setup of the variable multi-cusp magnetic field plasma device (MPD)*

Plasma's response to this perturbation is recorded in the electron saturation regime ($\delta I_{esat}$), which is detected at three different radial locations using a set of Langmuir probes placed at 2cm, 6cm, and 10cm, respectively, from the exciter disc. All three probes are in a uniform quiescence plasma region where all the plasma parameters are constant, and the magnetic field is minimum. Figure 2 shows the cross-section view of the device with exciter and probe array location. The perturbation is launched in the radial plane of the device, and a response is also recorded along the radial plane of the device. Each probe diameter is 1mm, and the tip length of 5mm. The probes are based near the local plasma potential ~12V to measure the Electron saturation current ($\delta I_{esat}$). These data are acquired using an 8-bit digital oscilloscope with different sampling rates and stored for further analysis.

## 3. Experimental results and Discussions

### 3.1 Soliton wave excitation and its characterization

H.Ikezi, R.J.Taylor, and D.R.Baker[2,56] have proved the ability of sinusoidal signals to produce soliton. Excitation of a soliton by a single sinusoidal pulse perturbation has not been extensively studied. In our experiment, emphasis has been given to exciting the wave by sinusoidal pulse, as described below.

Plasma's response to applied pulse sinusoidal wave perturbation is shown in figure 3 below. This perturbation signal shows in the top trace (red color) of figure 3. Response of these perturbation signals is recorded in electron saturation ($\delta I_{esat}$) regime by the probe array located at different locations of the device. The lower three signals are the probe signals. These signals are normalized with their own maxima, as shown in the figure; hence the Y-axis of all subfigures is on a -1 to 1 scale. The X-axis shows the temporal evolution.

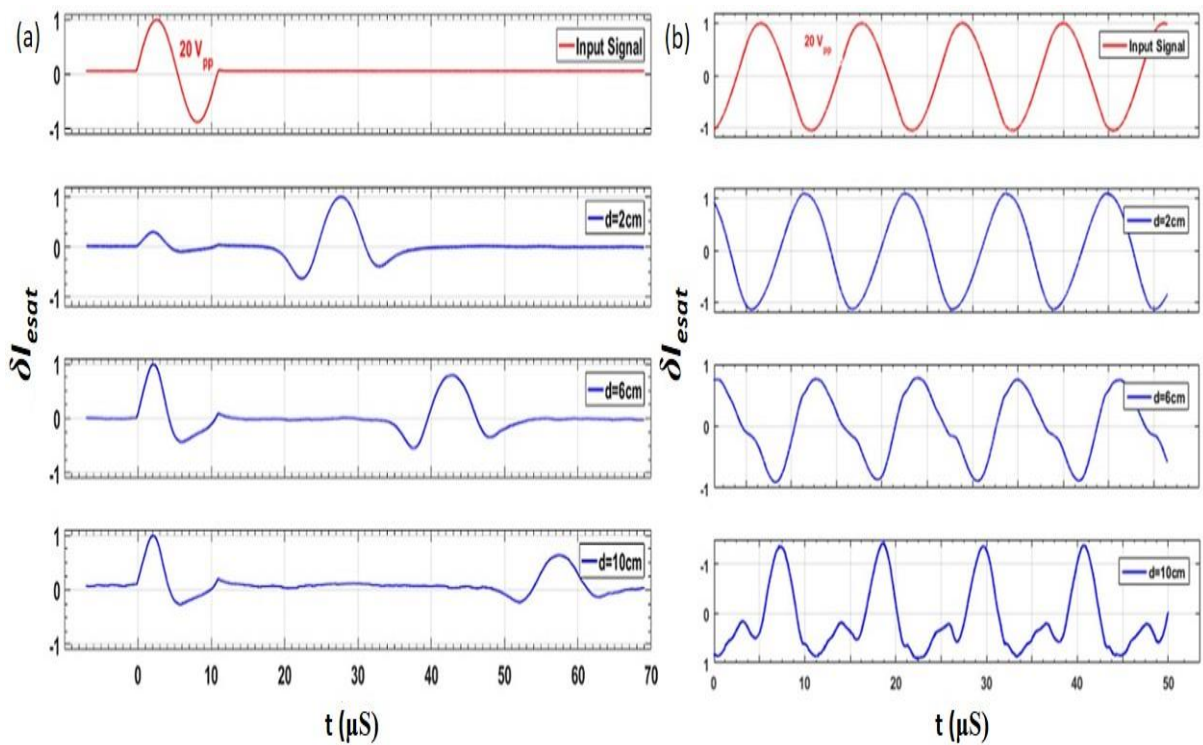

*Figure 3(a): Time evolution of single pulse sine wave perturbation recorded at the different radial locations*
*Figure 3(b): Time evolution of continuous sinusoidal wave perturbation recorded at different radial locations*

The time evolution of a single pulse sine wave has been shown in figure 3a. It is observed that there are two types of waveforms present in this received signal.

   A. **Ion Burst:** The first signal rises and falls with the time scale of the perturbation voltage. This waveform is reported or known as Ion burst or Ballistic mode[38,57,58]. This ion burst signal is a misleading addition unwanted signal. It was fetched in by the exciter disk. Sometimes, it is confused with fast electron response, capacitive pickup, or noise. The root cause of this signal is due to free streaming of ions. In this paper, we are not further perusing any study of this Ion burst mode. Our current primary focus is on later appearing time delayed waveform.

   B. **Ion Acoustic Soliton**:

   As this wave has been excited by the sinusoidal wave, both parts of the sine wave (i.e., positive and negative) are taking part in perturbation. Hence, in the received signal initially, it goes towards a negative direction, then it starts rising towards a positive direction and shows a shape relevant to the hyperbolic secant. From the observed time delay in the received signal, the velocity has been calculated by the time of flight method.

   The width and the velocity are compared as a function of its amplitude. To obtain this relation, different amplitude soliton was excited in plasma. As the amplitude of the soliton increases, the width of the wave decreases, and as amplitude decreases, the width increases.

   The soliton velocity is obtained by the time of flight technique by averaging the propagation in plasma. This wave is having velocity higher than the Ion acoustic velocity. This experimentally measured velocity is normalized with $C_s$ to obtain the Mach number. Figure 4 shows the Mach number plot. The experimentally measured velocity normalized to $C_s$ is plotted in the Blue line with diamond markers. The theoretical line (red line with star markers) is calculated from equation 3. The dependency of Mach number on its amplitude is compared with theory, and experimental are shown in figure 4.

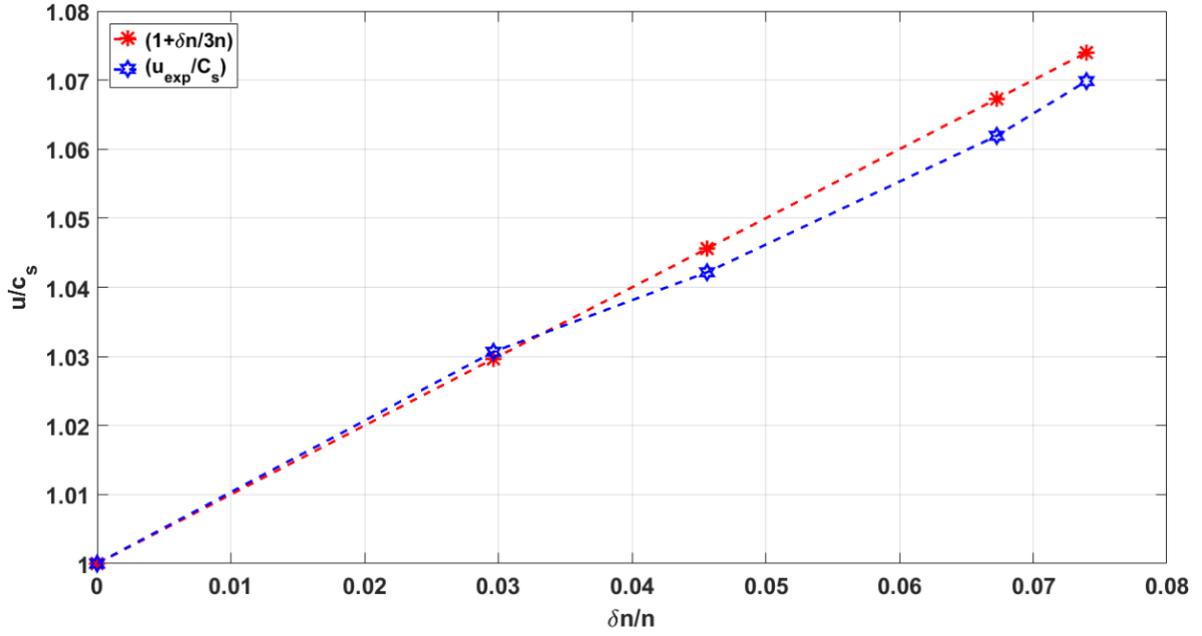

*Figure 4: Velocity of the soliton as a function of its amplitude δn*

The width of the soliton is measured experimentally as follows. First, the full width at half maximum was measured from the temporal evolution shown in figure 3a. This will give the temporal width. To obtain the spatial width, temporal width is multiplied by the obtained velocity of the soliton. This will give the soliton width D. the square of the width is normalized by $\lambda_D$. The unperturbed plasma density is normalized with the amplitude $\delta n$ and plotted in the X-axis of figure5.

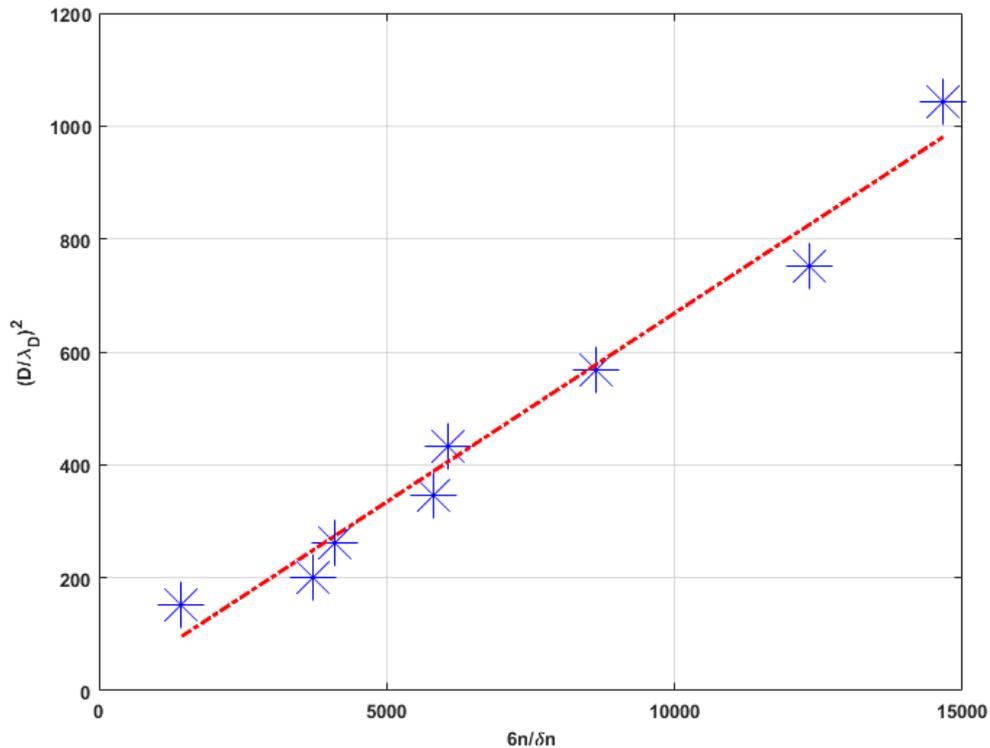

*Figure5: The width D of the soliton as a function of its amplitude δn.*

Figure 4 & 5 shows a good match with theory and experimental results, which agrees with the abovementioned KdV equation. These characteristics of the wave are reasonable to identify them as a soliton.

## 3.2 Interaction of two counter-propagating Soliton

The interaction of two solitons has been studied in MPD to substantiate that the propagating wave is a soliton. When two solitary waves collide, they overlap and pass through each other without losing their identity. The analysis of numerical solutions to the Korteweg-de Vries (KdV) equation also established this; such waves are known as a soliton. Slyunyaev[59] theoretically studied the interaction of solitons described by solutions to the KdV equation.

To excite the two counter-propagating solitons, another exciter disc of the same shape and size is placed on another side of the Langmuir probe set in plasma. These probes and exciters are in a uniform field-free region. The perturbation was applied to both exciters simultaneously. Perturbation voltage and amplitude were kept constant as described above, frequency $90 kH_z$ and amplitude $\sim 20 V_{pp}$. Figure 6 shows the interaction of two counter-propagating solitary waves.

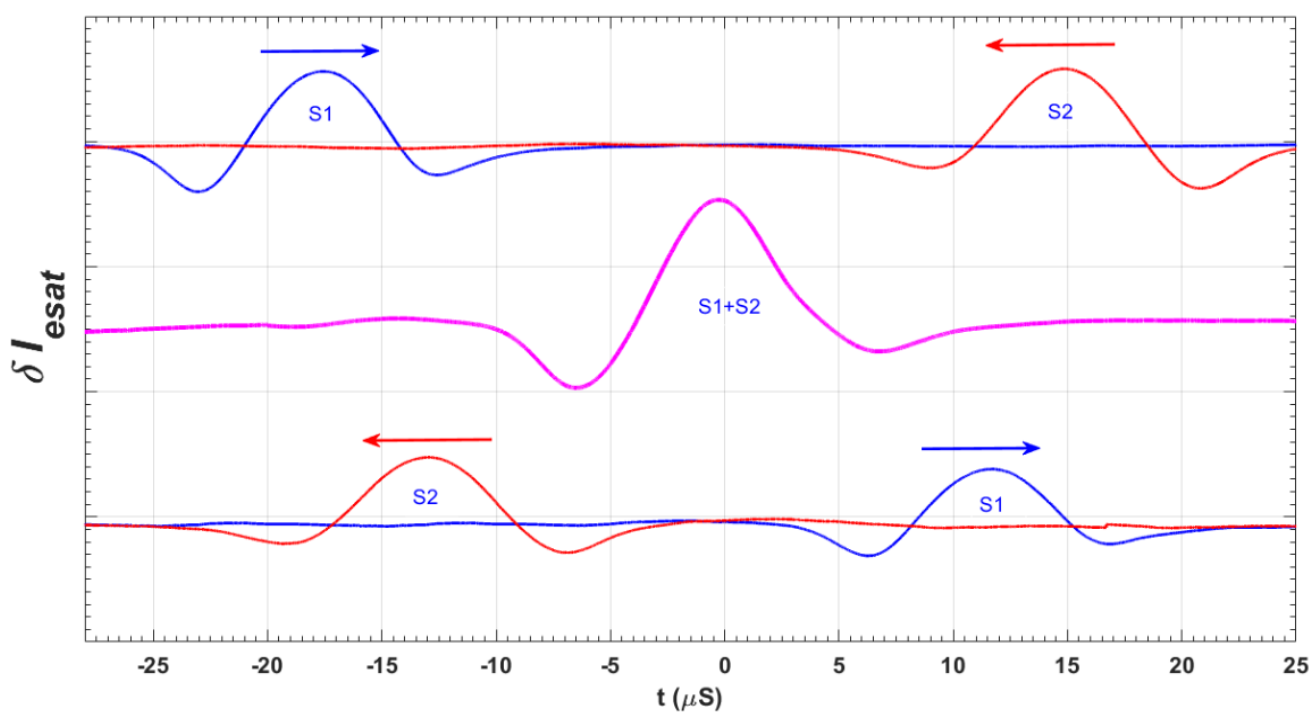

*Figure 6: Interaction between two counter-propagating solitons*

The top trace of the figure shows the two solitons, S1 and S2, are excited from the individual exciters and propagating towards each other. S1 and S2 have the same amplitude and velocity. S1 and S2 interact at the center of the two exciters, merge into each other linearly, and generate a single soliton, shown in the mid trace. After the interaction, they separate and travel ahead without losing their identity.

By above mentioned three characteristics, i.e., dependency of velocity and width on its amplitude and interaction between two counter-propagating solitons, definite it to be soliton.

## 4. Effect of Cusp Magnetic field on the Propagation of Ion acoustic Soliton

Earlier, people used permanent magnets to produce the cusp magnetic field[60]. In MPD, Electromagnets produce the cusp magnetic field, which gives freedom to change the cusp magnetic field strength by changing the applied currents to magnets. Ion Acoustic soliton is excited in the uniform field-free region where the ions are unmagnetized, and plasma is uniform and quiescent. The cusp magnetic field configuration provides exceptional macroscopic plasma consistency due to U-shaped magnetic field curvature towards the confine plasma system in the center, and plasma is also stable to large-scale perturbation. It is well known that the cusp magnetic field confines the maximum energy of the primary or hot energetic electrons[60,61,62]. In MPD, as the magnetic field increases, primary electrons' confinement also increases. The primary electron confinement reaches its maxima at $B_p \sim 0.6\ kG$ ($I_{mag} = 80A$) and remains constant as the magnetic field increases above this value. Beyond $B_p \sim 0.6 kG$, thermalization reduces the confinement of the primary hot energy electrons. Here $I_{mag} = 80A$ is the applied magnet current that passes through the electromagnets to produce the respective magnetic field of $B_p \sim 0.6 kG$ at the pole of the magnets. To observe the effect of the cusp magnetic field on soliton, the solitons were excited n a similar way described above, and the pole magnetic field was varied. Earlier, we found that with the increasing magnetic field, all the plasma parameters are uniform in the field-free region area[50].

Figure 9 shows the soliton amplitude and width with increasing pole cusp magnetic field values; as the magnetic field value is increased initially, soliton amplitude increase with a magnetic field. At $\sim 0.6 kG$ ($I_{mag} = 80A$), the amplitude has the maximum value. After this, as we increase the magnetic field, the soliton amplitude starts decreasing gradually. As the amplitude of the soliton increases, the width starts decreasing, and beyond $\sim 0.6 kG$ ($I_{mag} = $ 80$A$), it starts increasing gradually. The opposite variation of the soliton amplitude and width is evident from figure 9, that soliton is sustained in plasma at each applied magnetic field.

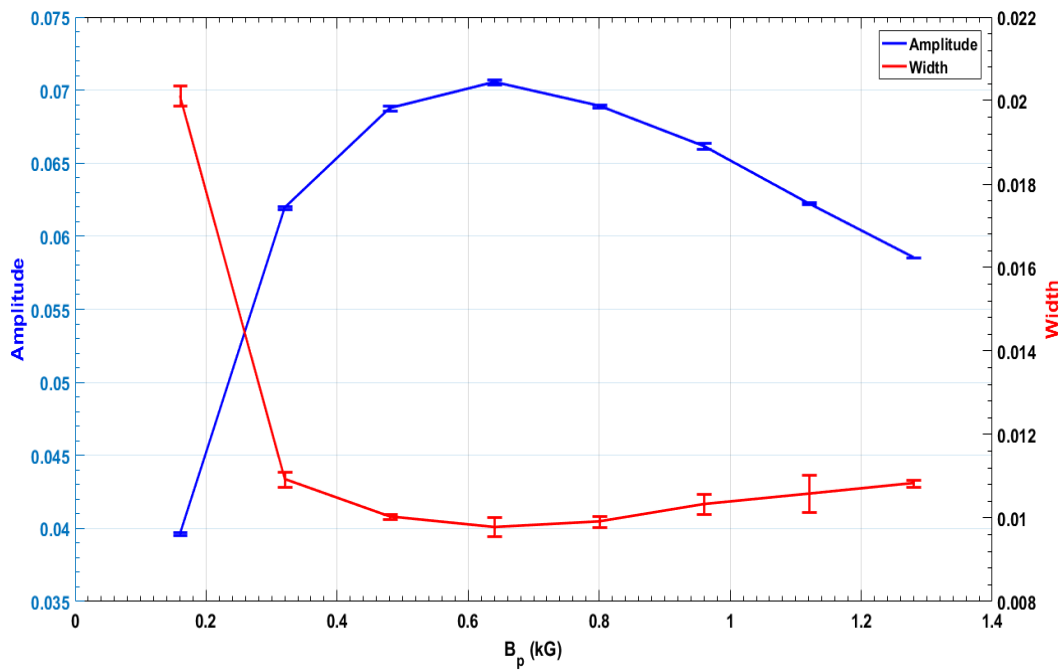

*Figure 7: Amplitude and Width of soliton with increasing multi-pole cusp magnetic field.*

When the exciter applies a potential perturbation in the plasma, electrons will react to shield this potential. When a positive potential is applied, they (electrons) will move towards exciter from all nearby locations. Since the shielding is not perfect due to temperature, the ions also feel the leaked potential. When they (ions) start moving, potential reverses, and the electrons are pushed away, and because of inertia, they overshoot. By this mechanism, the wave moves in plasma.

Figure 10 shows the relative hot electron density and floating potential variation with different magnetic field values. The left side of the Y-axis in figure 10 shows the Relative hot electron density $N_c/N_h$, and the right side of the Y-axis shows the floating potential ($V_f$) with different magnetic field values.

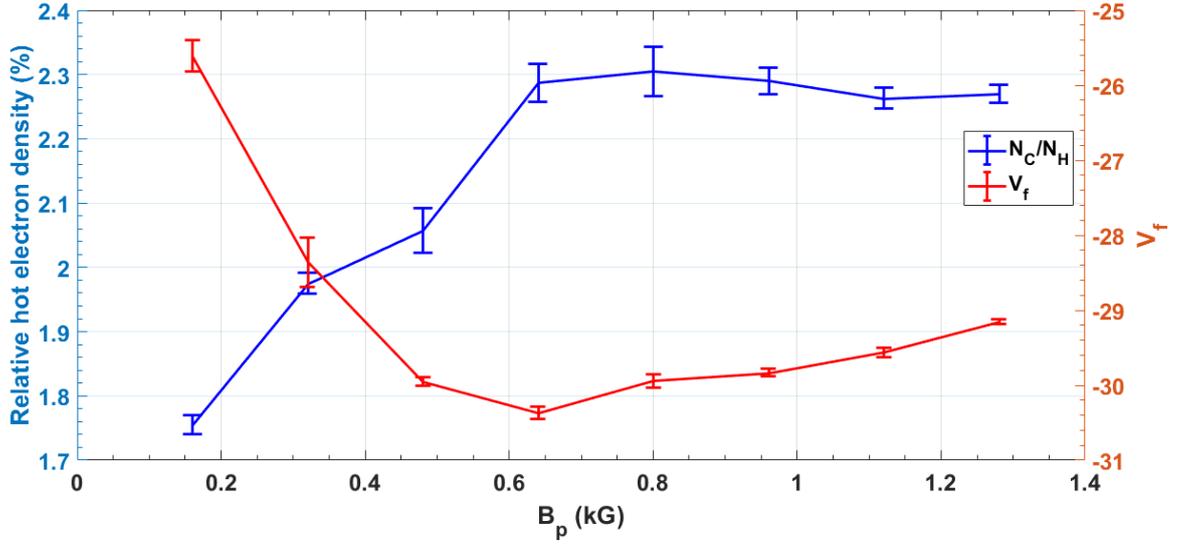

*Figure 8: Relative hot electron density and variation of floating potential with increasing cusp magnetic field*

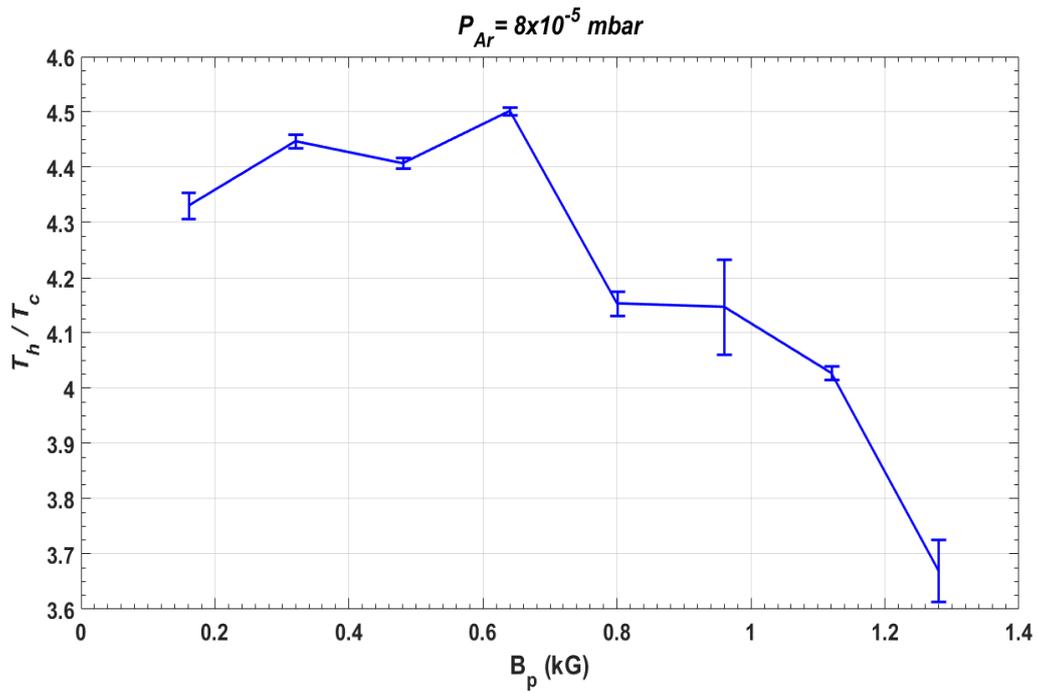

*Figure 9: Ratio of hot electron temperature to cold electron temperature with increasing multi-pole line cusp magnetic field*

As the magnetic field is varied up to $\sim 0.6 kG$ ($I_{mag} = 80A$), floating potential increases, decreases slightly afterward, and becomes saturated. This observation shows that the confinement of the primary electrons is maximum at $\sim 0.6 kG$ ($I_{mag} = 80A$). The soliton amplitude follows the same trend of floating potential.

The floating potential measurement indicates the population of primary electrons; hence the relative hot electron temperature and density show a similar trend with floating potential. Jones[63] et al. validated that the small fraction of cool electrons can dictate ion acoustic wave propagation. Figure 11 shows the hot electron temperature ($T_h$) ratio to cold electron temperature $T_c$ with an increasing multi-pole line cusp magnetic field. As the magnetic field increases, the primary electrons are getting confined, achieving maximum confinement at ~$0.6 kG$ ($I_{mag} = 80A$); beyond that ratio of $T_h/T_c$ start falling with the cusp magnetic field. $T_h/T_c$ also follows the similar trend of $V_f$ so does the amplitude of soliton.

## 5. Conclusion

In our studies, we have excited the soliton in MPD by sinusoidal perturbation and characterized them. The propagating wave satisfies the relation between the amplitude, the Mach number, and width of the solitary wave as described by the KdV equation. Two counter-propagating solitons were found to overlap and pass through each other without losing their identity. The maximum amplitude of the soliton generated in the present experiment is $B_p = 0.6 kG$ ($I_{mag} = 80A$); hence the excitation and interaction experiments were performed with $B_p = 0.6 kG$ pole cusp magnetic field.

It is experimentally observed in MPD that pole cusp field value influences the ion-acoustic soliton. The amplitude of solitons was found to increase with the field value $B_p = 0.6 kG$, then decreases with a further increase in field values. The width of the soliton shows the opposite variation to amplitude. The soliton evolution is found to be much sensitive to primary high energetic electron confinement by the cusp magnetic field as the amplitude and width of soliton vary significantly with it.

## Acknowledgements

The authors would like to thank Dr. Joydeep Ghosh for the critical review of the manuscript. Z.S is indebted to University Grant Commission (UGC) and Ministry of Minority affairs, Government of India for their support under the Maulana Azad National Fellowship scheme award letter number 2016-17/MANF-2015-17-GUJ-67921.